\documentclass[aps,prl,showpacs,reprint,superscriptaddress]{revtex4-1}
\usepackage{graphicx}
\usepackage{bm}
\usepackage{amsmath}

\begin{document}

\title{Quantum Computation with Rotational States of Nonpolar Ionic Molecules}

\author{Sang Jae Yun}
\email[]{sangjae@kaist.ac.kr}
\affiliation{Department of Physics, Korea Advanced Institute of Science and Technology, Daejeon 305-701, Korea}
\affiliation{Center for Relativistic Laser Science, Institute for Basic Science, Gwangju 500-712, Korea}
\author{Chang Hee Nam}
\affiliation{Department of Physics, Korea Advanced Institute of Science and Technology, Daejeon 305-701, Korea}
\affiliation{Center for Relativistic Laser Science, Institute for Basic Science, Gwangju 500-712, Korea}
\affiliation{Department of Physics and Photon Science, Gwangju Institute of Science and Technology, Gwangju 500-712, Korea}

\date{4 April 2013}

\begin{abstract}
We propose a quantum computer architecture which is robust against decoherence and scalable. As a qubit, we adopt rotational states of a nonpolar ionic molecule trapped in an ion-trap. It is revealed that the rotational-state qubits are much more immune to decoherence than the conventional electronic-state qubits of atomic ions. A complete method set for state preparation, single-qubit gate, controlled-NOT gate, and qubit-readout suitable for the rotational-state qubits is provided. Since the ionic molecules can be transported in an array of ion traps, the rotational-state qubits are expected to be a promising candidate to build a large-scale quantum computer. 
\end{abstract}

\pacs{03.67.Lx, 33.20.Sn, 37.10.Ty, 42.50.Ex}

\maketitle


A quantum computer (QC) is a computing machine that uses quantum logic totally different from the Boolean logic on which classical computers are built. If a QC is successfully built, it will make numerous unimaginable things possible, in a similar manner that lasers have done \cite{Ladd2010}. The most critical difficulty in building a QC is that qubits (two-level quantum systems) should be well isolated from environment to preserve their fragile quantum states from decoherence, while at the same time a quantum channel should exist to give programmers access to the qubits. Another serious difficulty, called scalability, arises from the fact that one should manipulate at least thousands of qubits to perform a practical computation \cite{Cirac2000}. Since all elementary building blocks necessary for construction of a QC have been demonstrated over the past decade, current goals intensively pursued nowadays are scaling up to a larger number of qubits and raising the fidelity of qubit-manipulation \cite{Benhelm2008}. In accomplishing these missions, the most critical obstacle is the decoherence problem. Thus, it is desirable to find out a new qubit system immune to decoherence.

We propose a novel ion-trap-based QC architecture that is robust against decoherence and scalable. As a qubit, we adopt rotational states of a nonpolar molecular ion trapped in an ion-trap. To our knowledge, although rotational states of neutral polar molecules were already proposed as qubits \cite{Demille2002, Andre2006}, those of ionic nonpolar molecules have not been proposed. Without collisions, the lifetime of rotational states of a nonpolar molecule can reach several years \cite{Lee2004} because the rotational state is electromagnetically inactive owing to no permanent dipole moment. We reveal that the rotational-state qubits of nonpolar molecules are much more immune to decoherence than the conventional electronic-state qubits of atomic ions, and provide a complete method set necessary for building a QC such as state preparation, single-qubit gate, controlled-NOT gate, and qubit-readout appropriate to the rotational-state qubits. Since we adopt ionic molecules, one can utilize well-developed ion-trap technologies, such as separation, transportation, and combination of individual ions, which are the key ingredients to implement a large-scale QC \cite{Kielpinski2002}. The movement control was already demonstrated for atomic ions \cite{Home2009}, and no additional restriction is imposed to molecular ions. Because nonpolar quantum rotors are excellent gyroscope, their rotational states should be very robust under the transportation. 

We first survey general features of rotational states of nonpolar linear molecules and select two levels as a qubit. Rotationally stationary states of a linear molecule are angular momentum eigenstates $\left| {J,M} \right\rangle $, where $J$ is the angular momentum quantum number and $M$ is its z-axis projection in the laboratory frame. Because a nonpolar molecule has no permanent dipole moment, laser field cannot cause dipole transitions ($\Delta J =  \pm 1$). However, since linear molecules have anisotropy in their polarizability, they allow rotational two-photon Raman transitions with $\Delta J =  \pm 2$. When the laser field is linearly polarized along the z-axis, one additional selection rule gives $\Delta M = 0$. With these selection rules, one can choose a qubit system as $\left|  \downarrow  \right\rangle  \equiv \left| {{J_0},{M_0}} \right\rangle $ and $\left|  \uparrow  \right\rangle  \equiv \left| {{J_0} + 2,{M_0}} \right\rangle $ for arbitrary ${J_0}$ and ${M_0}$. But choosing ${J_0}$ and ${M_0}$ is restricted according to molecular species, because nuclear spin statistics categorize nonpolar linear molecules into three groups: even-, odd-, and all-$J$ molecules \cite{Herzberg1989}. In order to be a nonpolar molecule, the nuclei at both ends of the molecule should be identical particles obeying the exchange symmetry of quantum mechanics. For example, in ${\rm{NS}}_{\rm{2}}^{\rm{ + }}$, the two ${}^{{\rm{32}}}{\rm{S}}$ are spin-zero bosons and the electronic ground state is ${}^1\Sigma _g^ + $ so that only even-$J$ states are allowed. Among the three types, even-$J$ is the most preferred because the rotational ground state of this kind is $\left| {0,0} \right\rangle $ which is nondegenerate. On the contrary, the ground state of an odd-$J$ molecule is degenerate and can be either of $\left| {1, - 1} \right\rangle $, $\left| {1, 0} \right\rangle $ or $\left| {1, 1} \right\rangle $ so that additional selection is needed to prepare a qubit state even at zero temperature. Moreover, the ground state of an all-$J$ molecule is more ambiguous because it depends on the spin states of the nuclei. In this letter, we assume that an even-$J$ molecule is chosen and the qubit system is $\left|  \downarrow  \right\rangle  \equiv \left| {0,0} \right\rangle $ and $\left|  \uparrow  \right\rangle  \equiv \left| {2,0} \right\rangle $. Because the Hilbert space of rotational states of an even-$J$ molecule intrinsically excludes odd-$J$ states, one additional merit arises that there is no possibility of state-leakage into unwanted odd-$J$ space. 

In the view point of decoherence, we compare the rotational-state qubits with conventional electronic-state qubits. In an ion-trap QC, the major source of decoherence is a fluctuating magnetic field \cite{Langer2005}. In the conventional atomic-ion architecture, qubits are made of electronic states of alkaline earth ions such as ${\rm{B}}{{\rm{e}}^ + }$ or ${\rm{M}}{{\rm{g}}^ + }$ that has only one open-shell electron. This makes electronic spin states a doublet with a big magnetic dipole moment, which results in a strong coupling between the qubit and the magnetic field through the Zeeman effect. On the contrary, in the case of the molecular rotational-state qubits, the magnetic moment of electrons affects the qubit state merely indirectly. Furthermore, there exist molecular ions having no electronic magnetic moment, such that the electronic ground state is ${}^1\Sigma $ state. In that case, the magnetic field affects the rotational-state qubit, only through the rotational magnetic moment ${\mu _r}$, which is proportional to the rotational g-factor ${g_r}$ and the rotational angular momentum $J$ as expressed by ${\mu _r} = {g_r}{\mu _N}\sqrt {J(J + 1)} $ \cite{Herzberg1989}, where ${\mu _N}$ is the nuclear magneton. Most nonpolar molecules have g-factors smaller than 1 \cite{Herzberg1989}, so ${\mu _r}$ is very small compared to the electronic spin magnetic moment which is equal to one Bohr magneton $1840{\mu _N}$ that the conventional electronic-state qubits have. It is thus expected that the rotational-state qubits are very robust against decoherence from magnetic field fluctuations. 

Several techniques have been adopted to improve the coherence time of atomic ions. For example, one can choose electronic-state qubits which are insensitive under magnetic fields to the first order \cite{Olmschenk2007, Langer2005} and can adopt the decoherence-free subspace (DFS) technique \cite{Kielpinski2001, Haffner2005}. With these techniques, single-qubit coherence time of an atomic ion has been achieved to be about 20 s \cite{Haffner2005}. But 20 s is not sufficient for a large-scale quantum computer, because, when the coherence time of a single qubit is $\tau $, that of $N$ qubits is approximately $\tau /N$ \cite{KielpinskiThesis}. The limited coherence time is mainly due to high-order Zeeman shift and local field fluctuation different at each ion composing a DFS pair. These two effects will be reduced if the magnetic moment of a qubit is small. Thus, the tiny magnetic moment of the rotational-state qubit can greatly extend the coherence time, several orders longer than the conventional electronic-state qubits of atomic ions. Furthermore, the rotational-state qubit is also first-order insensitive to the magnetic field because we chose the qubit states with $M = 0$, and the DFS technique can also be applied to rotational-state qubits. 

We examined a variety of molecular ions to choose a suitable candidate that satisfies the requirements including linearity, nonpolarity, even-$J$, electronic ${}^1\Sigma $ state, and small rotational g-factor. If we choose the candidate among singly-charged molecular ions, triatomic molecular ions, such as ${\rm{NO}}_{\rm{2}}^{\rm{ + }}$ or ${\rm{NS}}_{\rm{2}}^{\rm{ + }}$, can fulfill the requirements. Considering rotational g-factors calculated by the DALTON software \cite{Dalton2011}, we suggest ${\rm{NS}}_{\rm{2}}^{\rm{ + }}$ as a suitable candidate due to its tiny rotational g-factor of $-0.014$, which means that the magnetic moment of $\left|  \downarrow  \right\rangle $ is zero due to $J = 0$ and that of $\left|  \uparrow  \right\rangle $ is $ - 0.034{\mu _N}$. If we extend our choice to doubly-charged molecular ions, homonuclear diatomic dications can also be chosen. Although most diatomic dications are metastable, the lifetime of some species are very long; e.g., ${\rm{Ba}}_{\rm{2}}^{{\rm{2 + }}}$ has a lifetime longer than the age of the universe \cite{Sramek1980}. ${\rm{Ba}}_{\rm{2}}^{{\rm{2 + }}}$ not only fulfills the above requirements but also has an advantage of no nuclear spin. Because the nuclear spin causes the hyperfine spin-rotation coupling \cite{Amano2010JCP}, the nuclear spin can be another source of decoherence. While singly-charged triatomic ${}^1\Sigma $ molecular ions cannot have zero nuclear spin in their center nucleus, diatomic dications can simultaneously fulfill the two requirements, ${}^1\Sigma $ electronic state and zero nuclear spin. In this respect, very-long-lived homonuclear diatomic dications like ${\rm{Ba}}_{\rm{2}}^{{\rm{2 + }}}$ can also be a proper candidate. 

The first step in operating a QC is the state preparation of qubits, which should include translational, vibrational, and rotational cooling of molecular ions. The translational cooling of molecular ions can be achieved by the sympathetic cooling method \cite{Ostendorf2006} that relies on co-trapped atomic alkaline earth ions which can be easily cooled down using a laser. For cooling rovibrational states of molecules after the translational cooling, sophisticated techniques are needed, and its experimental demonstrations were reported only recently and moreover for only polar molecular ions \cite{Staanum2010, Schneider2010}. For nonpolar molecular ions, a novel method that exploits the motional ground state shared with atomic cooling-pair ions was proposed in 2012 \cite{Ding2012, Leibfried2012}. Although its experimental demonstration has not been reported yet, this method is expected to make nonpolar molecular ions reach the rotational ground state, i.e. $\left|  \downarrow  \right\rangle $, with high fidelity of more than 99.9\% in near future \cite{Leibfried2012}. 

A single-qubit operation can be performed through a resonant two-photon Raman transition between qubit levels. The Hamiltonian for rotational states of a nonpolar molecule interacting in a laser field is given by \cite{Bartels2001, Yun2012} 
\begin{equation}
{\bf{H}}(t) = {B_0}{{\bf{J}}^2} - {1 \over 2}\Delta \alpha {{\cal E}^2}(t){\cos ^2}\theta ,
\end{equation}
where ${B_0}$ is the molecular rotational constant, $\Delta \alpha  \equiv {\alpha _\parallel } - {\alpha _ \bot }$ is the anisotropy of molecular polarizability, ${\cal E}(t)$ the laser electric field, and $\theta $ the angle between laser polarization and molecular axis. Because the energy of $\left| {J,M} \right\rangle $ is ${B_0}J(J + 1)$, the energy gap between qubit levels is ${\omega _0} = 6{B_0}$. A pair of laser beams linearly polarized along the z-direction which have frequencies ${\omega _1}$ and ${\omega _2}$ with the frequency difference equal to ${\omega _0}$ can cause the resonant two-photon transition for achieving the single-qubit operation, as shown in Fig. 1. Here the laser beams are preferably IR lasers detuned far from vibrational or electronic transitions of the molecular ion. If the amplitudes of the two laser beams are equally ${{{{\cal E}_0}} \mathord{\left/ {\vphantom {{{{\cal E}_0}} 2}} \right. \kern-\nulldelimiterspace} 2}$, the synthesized field is given by
\begin{equation}
{\vec {\cal E}_{\left|  \downarrow  \right\rangle  \leftrightarrow \left|  \uparrow  \right\rangle }}(t) = {\bf{\hat z}}{{\cal E}_0}\cos {{{\omega _1} + {\omega _2}} \over 2}t\cos \left( {{{{\omega _0}} \over 2}t + \varphi } \right)
\end{equation}
and brings about the resonant Raman-Rabi oscillation with the Rabi frequency 
\begin{equation}
\Omega  = {1 \over {8\hbar }}\Delta \alpha {\cal E}_0^2\left\langle  \downarrow  \right|{\cos ^2}\theta \left|  \uparrow  \right\rangle  = {{0.2981} \over {8\hbar }}\Delta \alpha {\cal E}_0^2.
\end{equation}
To be specific, the properties of ${\rm{NS}}_{\rm{2}}^{\rm{ + }}$ was calculated from the GAUSSIAN software \cite{Gaussian} to obtain ${B_0} = 3.44$ GHz, leading to ${\omega _0} = 20.64$ GHz, and $\Delta \alpha  = 8.47$ ${{\rm{{\AA}}}^3}$. In order to achieve a Rabi frequency $\Omega$ of e.g. 1 MHz with this molecule, the field intensity (corresponding to ${\cal E}_0^2$) of $2.5 \times {10^6}$ ${\rm{W/c}}{{\rm{m}}^{\rm{2}}}$ is needed. It is desirable that the two laser beams are co-propagating in order not to change the motional state during the single-qubit operation since the photon recoil force exerted on ions becomes minimum in that condition \cite{Meekhof1996}. 
\begin{figure}[tb]
        \centering\includegraphics[width=5cm]{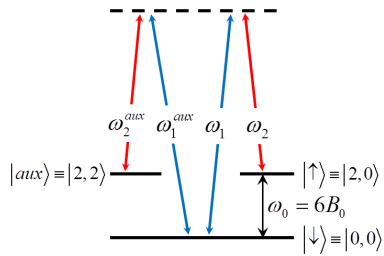}
        \caption{(Color online) Energy level diagram of qubit states. A pair of linearly-polarized laser beams with frequencies ${\omega _1}$ and ${\omega _2}$ induces a resonant two-photon Raman transition between the two qubit states for a single-qubit gate. The state $\left| {aux} \right\rangle $ and two circularly-polarized lasers with frequencies $\omega _1^{aux}$ and $\omega _2^{aux}$ are used when the Cirac-Zoller controlled-NOT gate is adopted.}
\end{figure}

The intermediate state in Fig. 1 is not a detuned real physical state, but a virtual state, because the interaction Hamiltonian of Eq. (1) induces the selection rule $\Delta J =  \pm 2$ which intrinsically requires two photons. This eliminates worry about detuning and state leakage into the intermediate state, which conventional electronic-hyperfine qubits of atomic ions suffer from. In this respect, the rotational-state qubits are expected to achieve a high fidelity and a high speed in gate operations owing to the no state leakage into the unwanted intermediate state. This is another advantage of the rotational-state qubits. 

In every proposed QC architecture, a key issue has been how a controlled-NOT gate is realized. Since any quantum logic gate can be decomposed into single-qubit operations and two-qubit controlled-NOT operations, the two kinds of operations constitute universal quantum gates \cite{Sleator1995}. A big advantage of the ion-trap-based QC is that the ion-trap provides an efficient way to carry out the controlled-NOT gate via collective motional states that the ions share. With the aid of the motional state, qubits can be effectively coupled even if their direct interaction is very weak. Several proposals have been made to realize the controlled-NOT gate in an ion trap \cite{Cirac1995, Cirac2000, Sorensen1999, Ospelkaus2008}, and some of them were demonstrated with electronic-state qubits of atomic ions \cite{SchmidtKaler2003, Lanyon2011, Ospelkaus2011}. Most of these proposals can also be applied to the rotational-state qubits of molecular ions as far as the ions are co-trapped in a trap potential. 

We consider two representative proposals and provide concrete methods to implement the proposals with the rotational-state qubits. Let us first consider the controlled-NOT gate proposed by Cirac and Zoller \cite{Cirac1995}. The Cirac-Zoller gate needs an auxiliary state $\left| {aux} \right\rangle $ coupled with $\left|  \downarrow  \right\rangle $ by another lasers without affecting $\left|  \uparrow  \right\rangle $. For the rotational-state qubit, we suggest $\left| {aux} \right\rangle  \equiv \left| {2,2} \right\rangle $ as shown in Fig. 1. The transition between $\left|  \downarrow  \right\rangle $ and $\left| {aux} \right\rangle $ can be performed by two circularly-polarized laser beams of which their polarization directions are counter-rotating in the x-y plane. If the two beams have frequencies $\omega _1^{aux}$ and $\omega _2^{aux}$ with frequency difference ${\omega _0}$ and their amplitudes are equally ${{{{\cal E}_0}} \mathord{\left/ {\vphantom {{{{\cal E}_0}} 2}} \right. \kern-\nulldelimiterspace} 2}$, the synthesized field is formed as 
\begin{gather}
{{\vec {\cal E}}_{\left|  \downarrow  \right\rangle  \leftrightarrow \left| {aux} \right\rangle }}(t) = {{\cal E}_0}\cos {{\omega _1^{aux} + \omega _2^{aux}} \over 2}t \nonumber
\\
 \times \left\{ {{\bf{\hat x}}\cos \left( {{{{\omega _0}} \over 2}t + \varphi } \right) + {\bf{\hat y}}\sin \left( {{{{\omega _0}} \over 2}t + \varphi } \right)} \right\}.
\end{gather}
This is a linearly-polarized field whose polarization direction is rotating in the x-y plane with a frequency of ${{{\omega _0}} \mathord{\left/
 {\vphantom {{{\omega _0}} 2}} \right.
 \kern-\nulldelimiterspace} 2}$, resulting in the selection rules $\Delta J =  \pm 2$ and $\Delta M =  \pm 2$. Owing to the selection rules, ${\vec {\cal E}_{\left|  \downarrow  \right\rangle  \leftrightarrow \left| {aux} \right\rangle }}$ causes the transition between $\left|  \downarrow  \right\rangle $ and $\left| {aux} \right\rangle $ without affecting $\left|  \uparrow  \right\rangle $. This exclusive transition is sufficient to implement the Cirac-Zoller gate. In contrast to the single-qubit gate, the Cirac-Zoller gate should generate the transition not only between $\left|  \downarrow  \right\rangle $ and $\left| {aux} \right\rangle $ but also between the collective motional states $\left| {n = 1} \right\rangle $ and $\left| {n = 0} \right\rangle $, where $n$ is the phonon quantum number. In order to alter the motional state, the actual frequency difference of the two laser beams is red-detuned such that ${\omega _0} \to {\omega _0} - \nu $, where $\nu $ is the phonon mode frequency along the z-axis. In addition to this detuning, if the laser beams are counter-propagating, the motional state is more efficiently changed than in the co-propagating condition \cite{Meekhof1996}. Additional requirements of this gate are that the motional state has to be cooled down to zero temperature and the laser beams should be addressed to one particular ion. 
 
These requirements were released in another controlled-NOT scheme proposed by S\o rensen and M\o lmer in 1999 \cite{Sorensen1999}. The S\o rensen-M\o lmer gate is performed through off-resonant laser pulses illuminating two ions simultaneously and does not need the auxiliary state. Because this method uses the collective motional state as a virtual intermediate state, the two-qubit gate works even in a motional thermal state. There is no restriction to adopt this method to the rotational-state qubits since the method works also in Raman transitions \cite{Sorensen1999}. 

\begin{figure}[tb]
        \centering\includegraphics[width=8cm]{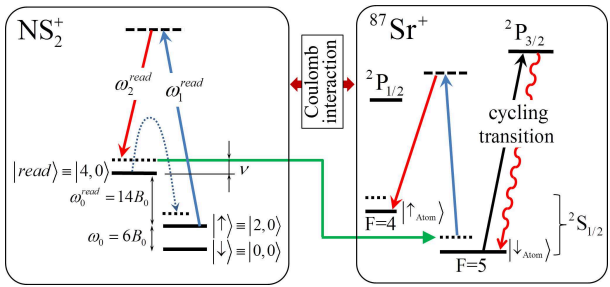}
        \caption{(Color online) Qubit readout scheme using state-transfer technique. Rotational-state qubits of molecular ions are readout by instead measuring electronic-state qubits of co-trapped atomic ions with fluorescence detection. Energy levels with dotted lines indicate the motional excited state $\left| {n = 1} \right\rangle $ with corresponding internal states.}
\end{figure}
At the end of computation, final qubit states should be read out. Previously, the state-of-the-art method of qubit readout has been based on fluorescence detection \cite{Blatt2008}. Unfortunately, the fluorescence-based method cannot be directly applied to the rotational-state qubits because, in the rotational states, there is no cycling transition decaying back to one of the qubit states. However, in an ion trap where two ions are co-trapped, an internal state of one ion can be transferred to that of the other ion \cite{Hume2007}. With this technique, the rotational-state qubit of a molecular ion can be readout by instead measuring the electronic-state qubit of an atomic ion that has a good cycling transition. Figure 2 illustrates this readout scheme. Alkaline earth ions such as ${}^{\rm{9}}{\rm{B}}{{\rm{e}}^{\rm{ + }}}$, ${}^{25}{\rm{M}}{{\rm{g}}^{\rm{ + }}}$, etc. have a good cycling transition and their hyperfine qubits can be measured with high fidelity. Among them, one having a similar mass with the molecular ion is preferred because the motional interaction between the two ions can be efficient \cite{Ding2012}. If one choose ${\rm{NS}}_{\rm{2}}^{\rm{ + }}$ molecule, we recommend ${}^{{\rm{87}}}{\rm{S}}{{\rm{r}}^{\rm{ + }}}$ due to their similar mass. The readout scheme consists of following 5 steps. (1) Sympathetically cooling down the shared motional state to zero temperature by laser cooling of the atomic ion. (2) State preparation of the atomic ion to $\left| {{ \downarrow _{{\rm{atom}}}}} \right\rangle $. (3) Blue-sideband Raman transition from $\left|  \uparrow  \right\rangle $ to $\left| {read} \right\rangle $ by applying a $\pi $-pulse detuned by $\nu $, where $\nu $ is the motional mode frequency, which means $\omega _1^{read} - \omega _2^{read} = \omega _0^{read} + \nu $. In this step, motional state is changed from $\left| {n = 0} \right\rangle $ to $\left| {n = 1} \right\rangle $ only if the rotational-state qubit was $\left|  \uparrow  \right\rangle $. The two Raman beams are preferred to counter-propagate along the motional axis, in order to efficiently change the motional quantum number \cite{Meekhof1996}. (4) Red-sideband Raman transition of the atomic ion from $\left| {{ \downarrow _{{\rm{atom}}}}} \right\rangle $ to $\left| {{ \uparrow _{{\rm{atom}}}}} \right\rangle $. Only when the motional state was excited to $\left| {n = 1} \right\rangle $, this red-sideband transition actually occurs. (5) State-dependent fluorescence detection of the atomic ion using the cycling transition. If fluorescence is detected, the atomic state is measured as $\left| {{ \downarrow _{{\rm{atom}}}}} \right\rangle $, and the rotational-state qubit is considered as measured as $\left|  \downarrow  \right\rangle $, otherwise $\left|  \uparrow  \right\rangle $.

In order to enhance the fidelity of readout, the above procedures can be repetitively performed, with slight modification in step (3). Because, at the second repetition, the rotational state would lie on $\left| {read} \right\rangle $ (not on $\left|  \uparrow  \right\rangle $) with motional ground state $\left| {n = 0} \right\rangle $, red-sideband Raman transition from $\left| {read} \right\rangle $ to $\left|  \uparrow  \right\rangle $ (see the dotted arrow in Fig. 2) will excite the motional state. This causes the same effect as the former procedures and can transfer quantum states to the atomic ion. This way, readout can be performed several times. In every step (3), the blue- and red-sideband transitions alter in turn. This repetitive readout scheme is almost the same as the experiment done by Hume et al. \cite{Hume2007}. The only difference is that, in our case, the life time of $\left| {read} \right\rangle $ is very long (years), so $\left| {read} \right\rangle $ does not decay back to $\left|  \uparrow  \right\rangle $. This long lifetime of rotational states of nonpolar molecules is rather advantageous to achieve high fidelity. In the experiment of Hume et al., there was about 1\% probability of state leakage during the decay process, limiting the fidelity. In our case, there is no worry about state leakage owing to no decay. In this respect, our proposed readout scheme is expected to achieve at least 99.94\% fidelity which Hume et al. already achieved.

In conclusion, we have proposed a novel QC architecture using rotational-state qubits of nonpolar ionic molecules. It was shown that the rotational-state qubits are very immune to decoherence and can achieve high-fidelity logic gates owing to no state leakage into an intermediate state during qubit manipulations. A complete method set necessary for building a QC suitable for the rotational-state qubits was provided. The proposed architecture is expected to be one of the promising candidates for implementing a practical QC. 

\begin{acknowledgments}
We thank Hai Woong Lee, Jaewan Kim, and Jaewook Ahn
for useful comments. This work was supported by the Ministry of Education, Science and Technology of Korea through the National Research Foundation and Institute for Basic Science.
\end{acknowledgments}

\bibliography{Bib_total}

\begin{thebibliography}{35}%
\makeatletter
\providecommand \@ifxundefined [1]{%
 \@ifx{#1\undefined}
}%
\providecommand \@ifnum [1]{%
 \ifnum #1\expandafter \@firstoftwo
 \else \expandafter \@secondoftwo
 \fi
}%
\providecommand \@ifx [1]{%
 \ifx #1\expandafter \@firstoftwo
 \else \expandafter \@secondoftwo
 \fi
}%
\providecommand \natexlab [1]{#1}%
\providecommand \enquote  [1]{``#1''}%
\providecommand \bibnamefont  [1]{#1}%
\providecommand \bibfnamefont [1]{#1}%
\providecommand \citenamefont [1]{#1}%
\providecommand \href@noop [0]{\@secondoftwo}%
\providecommand \href [0]{\begingroup \@sanitize@url \@href}%
\providecommand \@href[1]{\@@startlink{#1}\@@href}%
\providecommand \@@href[1]{\endgroup#1\@@endlink}%
\providecommand \@sanitize@url [0]{\catcode `\\12\catcode `\$12\catcode
  `\&12\catcode `\#12\catcode `\^12\catcode `\_12\catcode `\%12\relax}%
\providecommand \@@startlink[1]{}%
\providecommand \@@endlink[0]{}%
\providecommand \url  [0]{\begingroup\@sanitize@url \@url }%
\providecommand \@url [1]{\endgroup\@href {#1}{\urlprefix }}%
\providecommand \urlprefix  [0]{URL }%
\providecommand \Eprint [0]{\href }%
\providecommand \doibase [0]{http://dx.doi.org/}%
\providecommand \selectlanguage [0]{\@gobble}%
\providecommand \bibinfo  [0]{\@secondoftwo}%
\providecommand \bibfield  [0]{\@secondoftwo}%
\providecommand \translation [1]{[#1]}%
\providecommand \BibitemOpen [0]{}%
\providecommand \bibitemStop [0]{}%
\providecommand \bibitemNoStop [0]{.\EOS\space}%
\providecommand \EOS [0]{\spacefactor3000\relax}%
\providecommand \BibitemShut  [1]{\csname bibitem#1\endcsname}%
\let\auto@bib@innerbib\@empty
\bibitem [{\citenamefont {Ladd}\ \emph {et~al.}(2010)\citenamefont {Ladd},
  \citenamefont {Jelezko}, \citenamefont {Laflamme}, \citenamefont {Nakamura},
  \citenamefont {Monroe},\ and\ \citenamefont {O'Brien}}]{Ladd2010}%
  \BibitemOpen
  \bibfield  {author} {\bibinfo {author} {\bibfnamefont {T.~D.}\ \bibnamefont
  {Ladd}}, \bibinfo {author} {\bibfnamefont {F.}~\bibnamefont {Jelezko}},
  \bibinfo {author} {\bibfnamefont {R.}~\bibnamefont {Laflamme}}, \bibinfo
  {author} {\bibfnamefont {Y.}~\bibnamefont {Nakamura}}, \bibinfo {author}
  {\bibfnamefont {C.}~\bibnamefont {Monroe}}, \ and\ \bibinfo {author}
  {\bibfnamefont {J.~L.}\ \bibnamefont {O'Brien}},\ }\href@noop {} {\bibfield
  {journal} {\bibinfo  {journal} {Nature}\ }\textbf {\bibinfo {volume} {464}},\
  \bibinfo {pages} {45} (\bibinfo {year} {2010})}\BibitemShut {NoStop}%
\bibitem [{\citenamefont {Cirac}\ and\ \citenamefont
  {Zoller}(2000)}]{Cirac2000}%
  \BibitemOpen
  \bibfield  {author} {\bibinfo {author} {\bibfnamefont {J.~I.}\ \bibnamefont
  {Cirac}}\ and\ \bibinfo {author} {\bibfnamefont {P.}~\bibnamefont {Zoller}},\
  }\href@noop {} {\bibfield  {journal} {\bibinfo  {journal} {Nature}\ }\textbf
  {\bibinfo {volume} {404}},\ \bibinfo {pages} {579} (\bibinfo {year}
  {2000})}\BibitemShut {NoStop}%
\bibitem [{\citenamefont {Benhelm}\ \emph {et~al.}(2008)\citenamefont
  {Benhelm}, \citenamefont {Kirchmair}, \citenamefont {Roos},\ and\
  \citenamefont {Blatt}}]{Benhelm2008}%
  \BibitemOpen
  \bibfield  {author} {\bibinfo {author} {\bibfnamefont {J.}~\bibnamefont
  {Benhelm}}, \bibinfo {author} {\bibfnamefont {G.}~\bibnamefont {Kirchmair}},
  \bibinfo {author} {\bibfnamefont {C.~F.}\ \bibnamefont {Roos}}, \ and\
  \bibinfo {author} {\bibfnamefont {R.}~\bibnamefont {Blatt}},\ }\href@noop {}
  {\bibfield  {journal} {\bibinfo  {journal} {Nat. Phys.}\ }\textbf {\bibinfo
  {volume} {4}},\ \bibinfo {pages} {463} (\bibinfo {year} {2008})}\BibitemShut
  {NoStop}%
\bibitem [{\citenamefont {DeMille}(2002)}]{Demille2002}%
  \BibitemOpen
  \bibfield  {author} {\bibinfo {author} {\bibfnamefont {D.}~\bibnamefont
  {DeMille}},\ }\href {\doibase 10.1103/PhysRevLett.88.067901} {\bibfield
  {journal} {\bibinfo  {journal} {Phys. Rev. Lett.}\ }\textbf {\bibinfo
  {volume} {88}},\ \bibinfo {pages} {067901} (\bibinfo {year}
  {2002})}\BibitemShut {NoStop}%
\bibitem [{\citenamefont {Andre}\ \emph {et~al.}(2006)\citenamefont {Andre},
  \citenamefont {DeMille}, \citenamefont {Doyle}, \citenamefont {Lukin},
  \citenamefont {Maxwell}, \citenamefont {Rabl}, \citenamefont {Schoelkopf},\
  and\ \citenamefont {Zoller}}]{Andre2006}%
  \BibitemOpen
  \bibfield  {author} {\bibinfo {author} {\bibfnamefont {A.}~\bibnamefont
  {Andre}}, \bibinfo {author} {\bibfnamefont {D.}~\bibnamefont {DeMille}},
  \bibinfo {author} {\bibfnamefont {J.~M.}\ \bibnamefont {Doyle}}, \bibinfo
  {author} {\bibfnamefont {M.~D.}\ \bibnamefont {Lukin}}, \bibinfo {author}
  {\bibfnamefont {S.~E.}\ \bibnamefont {Maxwell}}, \bibinfo {author}
  {\bibfnamefont {P.}~\bibnamefont {Rabl}}, \bibinfo {author} {\bibfnamefont
  {R.~J.}\ \bibnamefont {Schoelkopf}}, \ and\ \bibinfo {author} {\bibfnamefont
  {P.}~\bibnamefont {Zoller}},\ }\href@noop {} {\bibfield  {journal} {\bibinfo
  {journal} {Nat. Phys.}\ }\textbf {\bibinfo {volume} {2}},\ \bibinfo {pages}
  {636} (\bibinfo {year} {2006})}\BibitemShut {NoStop}%
\bibitem [{\citenamefont {Lee}\ \emph {et~al.}(2004)\citenamefont {Lee},
  \citenamefont {Villeneuve}, \citenamefont {Corkum},\ and\ \citenamefont
  {Shapiro}}]{Lee2004}%
  \BibitemOpen
  \bibfield  {author} {\bibinfo {author} {\bibfnamefont {K.~F.}\ \bibnamefont
  {Lee}}, \bibinfo {author} {\bibfnamefont {D.~M.}\ \bibnamefont {Villeneuve}},
  \bibinfo {author} {\bibfnamefont {P.~B.}\ \bibnamefont {Corkum}}, \ and\
  \bibinfo {author} {\bibfnamefont {E.~A.}\ \bibnamefont {Shapiro}},\ }\href
  {\doibase 10.1103/PhysRevLett.93.233601} {\bibfield  {journal} {\bibinfo
  {journal} {Phys. Rev. Lett.}\ }\textbf {\bibinfo {volume} {93}},\ \bibinfo
  {pages} {233601} (\bibinfo {year} {2004})}\BibitemShut {NoStop}%
\bibitem [{\citenamefont {Kielpinski}\ \emph {et~al.}(2002)\citenamefont
  {Kielpinski}, \citenamefont {Monroe},\ and\ \citenamefont
  {Wineland}}]{Kielpinski2002}%
  \BibitemOpen
  \bibfield  {author} {\bibinfo {author} {\bibfnamefont {D.}~\bibnamefont
  {Kielpinski}}, \bibinfo {author} {\bibfnamefont {C.}~\bibnamefont {Monroe}},
  \ and\ \bibinfo {author} {\bibfnamefont {D.~J.}\ \bibnamefont {Wineland}},\
  }\href@noop {} {\bibfield  {journal} {\bibinfo  {journal} {Nature}\ }\textbf
  {\bibinfo {volume} {417}},\ \bibinfo {pages} {709} (\bibinfo {year}
  {2002})}\BibitemShut {NoStop}%
\bibitem [{\citenamefont {Home}\ \emph {et~al.}(2009)\citenamefont {Home},
  \citenamefont {Hanneke}, \citenamefont {Jost}, \citenamefont {Amini},
  \citenamefont {Leibfried},\ and\ \citenamefont {Wineland}}]{Home2009}%
  \BibitemOpen
  \bibfield  {author} {\bibinfo {author} {\bibfnamefont {J.~P.}\ \bibnamefont
  {Home}}, \bibinfo {author} {\bibfnamefont {D.}~\bibnamefont {Hanneke}},
  \bibinfo {author} {\bibfnamefont {J.~D.}\ \bibnamefont {Jost}}, \bibinfo
  {author} {\bibfnamefont {J.~M.}\ \bibnamefont {Amini}}, \bibinfo {author}
  {\bibfnamefont {D.}~\bibnamefont {Leibfried}}, \ and\ \bibinfo {author}
  {\bibfnamefont {D.~J.}\ \bibnamefont {Wineland}},\ }\href {\doibase
  10.1126/science.1177077} {\bibfield  {journal} {\bibinfo  {journal}
  {Science}\ }\textbf {\bibinfo {volume} {325}},\ \bibinfo {pages} {1227}
  (\bibinfo {year} {2009})}\BibitemShut {NoStop}%
\bibitem [{\citenamefont {Herzberg}(1989)}]{Herzberg1989}%
  \BibitemOpen
  \bibfield  {author} {\bibinfo {author} {\bibfnamefont {G.}~\bibnamefont
  {Herzberg}},\ }\href {http://amazon.com/o/ASIN/0894642685/} {\emph {\bibinfo
  {title} {Molecular Spectra and Molecular Structure: Spectra of Diatomic
  Molecules}}},\ \bibinfo {edition} {2nd}\ ed.\ (\bibinfo  {publisher} {Krieger
  Pub Co, Malabar, Florida},\ \bibinfo {year} {1989})\BibitemShut {NoStop}%
\bibitem [{\citenamefont {Langer}\ \emph {et~al.}(2005)\citenamefont {Langer},
  \citenamefont {Ozeri}, \citenamefont {Jost}, \citenamefont {Chiaverini},
  \citenamefont {DeMarco}, \citenamefont {Ben-Kish}, \citenamefont {Blakestad},
  \citenamefont {Britton}, \citenamefont {Hume}, \citenamefont {Itano},
  \citenamefont {Leibfried}, \citenamefont {Reichle}, \citenamefont
  {Rosenband}, \citenamefont {Schaetz}, \citenamefont {Schmidt},\ and\
  \citenamefont {Wineland}}]{Langer2005}%
  \BibitemOpen
  \bibfield  {author} {\bibinfo {author} {\bibfnamefont {C.}~\bibnamefont
  {Langer}}, \bibinfo {author} {\bibfnamefont {R.}~\bibnamefont {Ozeri}},
  \bibinfo {author} {\bibfnamefont {J.~D.}\ \bibnamefont {Jost}}, \bibinfo
  {author} {\bibfnamefont {J.}~\bibnamefont {Chiaverini}}, \bibinfo {author}
  {\bibfnamefont {B.}~\bibnamefont {DeMarco}}, \bibinfo {author} {\bibfnamefont
  {A.}~\bibnamefont {Ben-Kish}}, \bibinfo {author} {\bibfnamefont {R.~B.}\
  \bibnamefont {Blakestad}}, \bibinfo {author} {\bibfnamefont {J.}~\bibnamefont
  {Britton}}, \bibinfo {author} {\bibfnamefont {D.~B.}\ \bibnamefont {Hume}},
  \bibinfo {author} {\bibfnamefont {W.~M.}\ \bibnamefont {Itano}}, \bibinfo
  {author} {\bibfnamefont {D.}~\bibnamefont {Leibfried}}, \bibinfo {author}
  {\bibfnamefont {R.}~\bibnamefont {Reichle}}, \bibinfo {author} {\bibfnamefont
  {T.}~\bibnamefont {Rosenband}}, \bibinfo {author} {\bibfnamefont
  {T.}~\bibnamefont {Schaetz}}, \bibinfo {author} {\bibfnamefont {P.~O.}\
  \bibnamefont {Schmidt}}, \ and\ \bibinfo {author} {\bibfnamefont {D.~J.}\
  \bibnamefont {Wineland}},\ }\href {\doibase 10.1103/PhysRevLett.95.060502}
  {\bibfield  {journal} {\bibinfo  {journal} {Phys. Rev. Lett.}\ }\textbf
  {\bibinfo {volume} {95}},\ \bibinfo {pages} {060502} (\bibinfo {year}
  {2005})}\BibitemShut {NoStop}%
\bibitem [{\citenamefont {Olmschenk}\ \emph {et~al.}(2007)\citenamefont
  {Olmschenk}, \citenamefont {Younge}, \citenamefont {Moehring}, \citenamefont
  {Matsukevich}, \citenamefont {Maunz},\ and\ \citenamefont
  {Monroe}}]{Olmschenk2007}%
  \BibitemOpen
  \bibfield  {author} {\bibinfo {author} {\bibfnamefont {S.}~\bibnamefont
  {Olmschenk}}, \bibinfo {author} {\bibfnamefont {K.~C.}\ \bibnamefont
  {Younge}}, \bibinfo {author} {\bibfnamefont {D.~L.}\ \bibnamefont
  {Moehring}}, \bibinfo {author} {\bibfnamefont {D.~N.}\ \bibnamefont
  {Matsukevich}}, \bibinfo {author} {\bibfnamefont {P.}~\bibnamefont {Maunz}},
  \ and\ \bibinfo {author} {\bibfnamefont {C.}~\bibnamefont {Monroe}},\ }\href
  {\doibase 10.1103/PhysRevA.76.052314} {\bibfield  {journal} {\bibinfo
  {journal} {Phys. Rev. A}\ }\textbf {\bibinfo {volume} {76}},\ \bibinfo
  {pages} {052314} (\bibinfo {year} {2007})}\BibitemShut {NoStop}%
\bibitem [{\citenamefont {Kielpinski}\ \emph {et~al.}(2001)\citenamefont
  {Kielpinski}, \citenamefont {Meyer}, \citenamefont {Rowe}, \citenamefont
  {Sackett}, \citenamefont {Itano}, \citenamefont {Monroe},\ and\ \citenamefont
  {Wineland}}]{Kielpinski2001}%
  \BibitemOpen
  \bibfield  {author} {\bibinfo {author} {\bibfnamefont {D.}~\bibnamefont
  {Kielpinski}}, \bibinfo {author} {\bibfnamefont {V.}~\bibnamefont {Meyer}},
  \bibinfo {author} {\bibfnamefont {M.~A.}\ \bibnamefont {Rowe}}, \bibinfo
  {author} {\bibfnamefont {C.~A.}\ \bibnamefont {Sackett}}, \bibinfo {author}
  {\bibfnamefont {W.~M.}\ \bibnamefont {Itano}}, \bibinfo {author}
  {\bibfnamefont {C.}~\bibnamefont {Monroe}}, \ and\ \bibinfo {author}
  {\bibfnamefont {D.~J.}\ \bibnamefont {Wineland}},\ }\href {\doibase
  10.1126/science.1057357} {\bibfield  {journal} {\bibinfo  {journal}
  {Science}\ }\textbf {\bibinfo {volume} {291}},\ \bibinfo {pages} {1013}
  (\bibinfo {year} {2001})}\BibitemShut {NoStop}%
\bibitem [{\citenamefont {H\"affner}\ \emph {et~al.}(2005)\citenamefont
  {H\"affner}, \citenamefont {Schmidt-Kaler}, \citenamefont {H\"ansel},
  \citenamefont {Roos}, \citenamefont {K\"orber}, \citenamefont {Chwalla},
  \citenamefont {Riebe}, \citenamefont {Benhelm}, \citenamefont {Rapol},
  \citenamefont {Becher},\ and\ \citenamefont {Blatt}}]{Haffner2005}%
  \BibitemOpen
  \bibfield  {author} {\bibinfo {author} {\bibfnamefont {H.}~\bibnamefont
  {H\"affner}}, \bibinfo {author} {\bibfnamefont {F.}~\bibnamefont
  {Schmidt-Kaler}}, \bibinfo {author} {\bibfnamefont {W.}~\bibnamefont
  {H\"ansel}}, \bibinfo {author} {\bibfnamefont {C.}~\bibnamefont {Roos}},
  \bibinfo {author} {\bibfnamefont {T.}~\bibnamefont {K\"orber}}, \bibinfo
  {author} {\bibfnamefont {M.}~\bibnamefont {Chwalla}}, \bibinfo {author}
  {\bibfnamefont {M.}~\bibnamefont {Riebe}}, \bibinfo {author} {\bibfnamefont
  {J.}~\bibnamefont {Benhelm}}, \bibinfo {author} {\bibfnamefont
  {U.}~\bibnamefont {Rapol}}, \bibinfo {author} {\bibfnamefont
  {C.}~\bibnamefont {Becher}}, \ and\ \bibinfo {author} {\bibfnamefont
  {R.}~\bibnamefont {Blatt}},\ }\href {\doibase 10.1007/s00340-005-1917-z}
  {\bibfield  {journal} {\bibinfo  {journal} {Applied Physics B}\ }\textbf
  {\bibinfo {volume} {81}},\ \bibinfo {pages} {151} (\bibinfo {year}
  {2005})}\BibitemShut {NoStop}%
\bibitem [{\citenamefont {Kielpinski}(2001)}]{KielpinskiThesis}%
  \BibitemOpen
  \bibfield  {author} {\bibinfo {author} {\bibfnamefont {D.}~\bibnamefont
  {Kielpinski}},\ }\emph {\bibinfo {title} {Entanglement and Decoherence in a
  Trapped-ion Quantum Register}},\ \href@noop {} {Ph.D. thesis},\ \bibinfo
  {school} {Univ. Colorado} (\bibinfo {year} {2001})\BibitemShut {NoStop}%
\bibitem [{Dal()}]{Dalton2011}%
  \BibitemOpen
  \href@noop {} {}\bibinfo {howpublished} {DALTON, a molecular electronic
  structure program, Release Dalton2011 (2011), see
  http://daltonprogram.org/}\BibitemShut {NoStop}%
\bibitem [{\citenamefont {Sramek}\ \emph {et~al.}(1980)\citenamefont {Sramek},
  \citenamefont {Macek},\ and\ \citenamefont {Gallup}}]{Sramek1980}%
  \BibitemOpen
  \bibfield  {author} {\bibinfo {author} {\bibfnamefont {S.~J.}\ \bibnamefont
  {Sramek}}, \bibinfo {author} {\bibfnamefont {J.~H.}\ \bibnamefont {Macek}}, \
  and\ \bibinfo {author} {\bibfnamefont {G.~A.}\ \bibnamefont {Gallup}},\
  }\href {\doibase 10.1103/PhysRevA.21.1361} {\bibfield  {journal} {\bibinfo
  {journal} {Phys. Rev. A}\ }\textbf {\bibinfo {volume} {21}},\ \bibinfo
  {pages} {1361} (\bibinfo {year} {1980})}\BibitemShut {NoStop}%
\bibitem [{\citenamefont {Amano}(2010)}]{Amano2010JCP}%
  \BibitemOpen
  \bibfield  {author} {\bibinfo {author} {\bibfnamefont {T.}~\bibnamefont
  {Amano}},\ }\href {\doibase 10.1063/1.3514914} {\bibfield  {journal}
  {\bibinfo  {journal} {J. Chem. Phys.}\ }\textbf {\bibinfo {volume} {133}},\
  \bibinfo {eid} {244305} (\bibinfo {year} {2010})}\BibitemShut {NoStop}%
\bibitem [{\citenamefont {Ostendorf}\ \emph {et~al.}(2006)\citenamefont
  {Ostendorf}, \citenamefont {Zhang}, \citenamefont {Wilson}, \citenamefont
  {Offenberg}, \citenamefont {Roth},\ and\ \citenamefont
  {Schiller}}]{Ostendorf2006}%
  \BibitemOpen
  \bibfield  {author} {\bibinfo {author} {\bibfnamefont {A.}~\bibnamefont
  {Ostendorf}}, \bibinfo {author} {\bibfnamefont {C.~B.}\ \bibnamefont
  {Zhang}}, \bibinfo {author} {\bibfnamefont {M.~A.}\ \bibnamefont {Wilson}},
  \bibinfo {author} {\bibfnamefont {D.}~\bibnamefont {Offenberg}}, \bibinfo
  {author} {\bibfnamefont {B.}~\bibnamefont {Roth}}, \ and\ \bibinfo {author}
  {\bibfnamefont {S.}~\bibnamefont {Schiller}},\ }\href {\doibase
  10.1103/PhysRevLett.97.243005} {\bibfield  {journal} {\bibinfo  {journal}
  {Phys. Rev. Lett.}\ }\textbf {\bibinfo {volume} {97}},\ \bibinfo {pages}
  {243005} (\bibinfo {year} {2006})}\BibitemShut {NoStop}%
\bibitem [{\citenamefont {Staanum}\ \emph {et~al.}(2010)\citenamefont
  {Staanum}, \citenamefont {Hojbjerre}, \citenamefont {Skyt}, \citenamefont
  {Hansen},\ and\ \citenamefont {Drewsen}}]{Staanum2010}%
  \BibitemOpen
  \bibfield  {author} {\bibinfo {author} {\bibfnamefont {P.~F.}\ \bibnamefont
  {Staanum}}, \bibinfo {author} {\bibfnamefont {K.}~\bibnamefont {Hojbjerre}},
  \bibinfo {author} {\bibfnamefont {P.~S.}\ \bibnamefont {Skyt}}, \bibinfo
  {author} {\bibfnamefont {A.~K.}\ \bibnamefont {Hansen}}, \ and\ \bibinfo
  {author} {\bibfnamefont {M.}~\bibnamefont {Drewsen}},\ }\href@noop {}
  {\bibfield  {journal} {\bibinfo  {journal} {Nat. Phys.}\ }\textbf {\bibinfo
  {volume} {6}},\ \bibinfo {pages} {271} (\bibinfo {year} {2010})}\BibitemShut
  {NoStop}%
\bibitem [{\citenamefont {Schneider}\ \emph {et~al.}(2010)\citenamefont
  {Schneider}, \citenamefont {Roth}, \citenamefont {Duncker}, \citenamefont
  {Ernsting},\ and\ \citenamefont {Schiller}}]{Schneider2010}%
  \BibitemOpen
  \bibfield  {author} {\bibinfo {author} {\bibfnamefont {T.}~\bibnamefont
  {Schneider}}, \bibinfo {author} {\bibfnamefont {B.}~\bibnamefont {Roth}},
  \bibinfo {author} {\bibfnamefont {H.}~\bibnamefont {Duncker}}, \bibinfo
  {author} {\bibfnamefont {I.}~\bibnamefont {Ernsting}}, \ and\ \bibinfo
  {author} {\bibfnamefont {S.}~\bibnamefont {Schiller}},\ }\href@noop {}
  {\bibfield  {journal} {\bibinfo  {journal} {Nat. Phys.}\ }\textbf {\bibinfo
  {volume} {6}},\ \bibinfo {pages} {275} (\bibinfo {year} {2010})}\BibitemShut
  {NoStop}%
\bibitem [{\citenamefont {Ding}\ and\ \citenamefont
  {Matsukevich}(2012)}]{Ding2012}%
  \BibitemOpen
  \bibfield  {author} {\bibinfo {author} {\bibfnamefont {S.}~\bibnamefont
  {Ding}}\ and\ \bibinfo {author} {\bibfnamefont {D.~N.}\ \bibnamefont
  {Matsukevich}},\ }\href@noop {} {\bibfield  {journal} {\bibinfo  {journal}
  {New J. Phys.}\ }\textbf {\bibinfo {volume} {14}},\ \bibinfo {pages} {023028}
  (\bibinfo {year} {2012})}\BibitemShut {NoStop}%
\bibitem [{\citenamefont {Leibfried}(2012)}]{Leibfried2012}%
  \BibitemOpen
  \bibfield  {author} {\bibinfo {author} {\bibfnamefont {D.}~\bibnamefont
  {Leibfried}},\ }\href@noop {} {\bibfield  {journal} {\bibinfo  {journal} {New
  J. Phys.}\ }\textbf {\bibinfo {volume} {14}},\ \bibinfo {pages} {023029}
  (\bibinfo {year} {2012})}\BibitemShut {NoStop}%
\bibitem [{\citenamefont {Bartels}\ \emph {et~al.}(2001)\citenamefont
  {Bartels}, \citenamefont {Weinacht}, \citenamefont {Wagner}, \citenamefont
  {Baertschy}, \citenamefont {Greene}, \citenamefont {Murnane},\ and\
  \citenamefont {Kapteyn}}]{Bartels2001}%
  \BibitemOpen
  \bibfield  {author} {\bibinfo {author} {\bibfnamefont {R.~A.}\ \bibnamefont
  {Bartels}}, \bibinfo {author} {\bibfnamefont {T.~C.}\ \bibnamefont
  {Weinacht}}, \bibinfo {author} {\bibfnamefont {N.}~\bibnamefont {Wagner}},
  \bibinfo {author} {\bibfnamefont {M.}~\bibnamefont {Baertschy}}, \bibinfo
  {author} {\bibfnamefont {C.~H.}\ \bibnamefont {Greene}}, \bibinfo {author}
  {\bibfnamefont {M.~M.}\ \bibnamefont {Murnane}}, \ and\ \bibinfo {author}
  {\bibfnamefont {H.~C.}\ \bibnamefont {Kapteyn}},\ }\href@noop {} {\bibfield
  {journal} {\bibinfo  {journal} {Phys. Rev. Lett.}\ }\textbf {\bibinfo
  {volume} {88}},\ \bibinfo {pages} {013903} (\bibinfo {year}
  {2001})}\BibitemShut {NoStop}%
\bibitem [{\citenamefont {Yun}\ \emph {et~al.}(2012)\citenamefont {Yun},
  \citenamefont {Kim}, \citenamefont {Lee},\ and\ \citenamefont
  {Nam}}]{Yun2012}%
  \BibitemOpen
  \bibfield  {author} {\bibinfo {author} {\bibfnamefont {S.~J.}\ \bibnamefont
  {Yun}}, \bibinfo {author} {\bibfnamefont {C.~M.}\ \bibnamefont {Kim}},
  \bibinfo {author} {\bibfnamefont {J.}~\bibnamefont {Lee}}, \ and\ \bibinfo
  {author} {\bibfnamefont {C.~H.}\ \bibnamefont {Nam}},\ }\href {\doibase
  10.1103/PhysRevA.86.051401} {\bibfield  {journal} {\bibinfo  {journal} {Phys.
  Rev. A}\ }\textbf {\bibinfo {volume} {86}},\ \bibinfo {pages} {051401}
  (\bibinfo {year} {2012})}\BibitemShut {NoStop}%
\bibitem [{\citenamefont {Frisch}\ \emph {et~al.}()\citenamefont {Frisch},
  \citenamefont {Trucks},\ and\ \citenamefont {\textit{et al.}}}]{Gaussian}%
  \BibitemOpen
  \bibfield  {author} {\bibinfo {author} {\bibfnamefont {M.~J.}\ \bibnamefont
  {Frisch}}, \bibinfo {author} {\bibfnamefont {G.~W.}\ \bibnamefont {Trucks}},
  \ and\ \bibinfo {author} {\bibfnamefont {H.~B.~S.}\ \bibnamefont {\textit{et
  al.}}},\ }\href@noop {} {}\bibinfo {note} {Gaussian~09 {R}evision {C}.01,
  Gaussian Inc. Wallingford CT 2010}\BibitemShut {NoStop}%
\bibitem [{\citenamefont {Meekhof}\ \emph {et~al.}(1996)\citenamefont
  {Meekhof}, \citenamefont {Monroe}, \citenamefont {King}, \citenamefont
  {Itano},\ and\ \citenamefont {Wineland}}]{Meekhof1996}%
  \BibitemOpen
  \bibfield  {author} {\bibinfo {author} {\bibfnamefont {D.~M.}\ \bibnamefont
  {Meekhof}}, \bibinfo {author} {\bibfnamefont {C.}~\bibnamefont {Monroe}},
  \bibinfo {author} {\bibfnamefont {B.~E.}\ \bibnamefont {King}}, \bibinfo
  {author} {\bibfnamefont {W.~M.}\ \bibnamefont {Itano}}, \ and\ \bibinfo
  {author} {\bibfnamefont {D.~J.}\ \bibnamefont {Wineland}},\ }\href {\doibase
  10.1103/PhysRevLett.76.1796} {\bibfield  {journal} {\bibinfo  {journal}
  {Phys. Rev. Lett.}\ }\textbf {\bibinfo {volume} {76}},\ \bibinfo {pages}
  {1796} (\bibinfo {year} {1996})}\BibitemShut {NoStop}%
\bibitem [{\citenamefont {Sleator}\ and\ \citenamefont
  {Weinfurter}(1995)}]{Sleator1995}%
  \BibitemOpen
  \bibfield  {author} {\bibinfo {author} {\bibfnamefont {T.}~\bibnamefont
  {Sleator}}\ and\ \bibinfo {author} {\bibfnamefont {H.}~\bibnamefont
  {Weinfurter}},\ }\href {\doibase 10.1103/PhysRevLett.74.4087} {\bibfield
  {journal} {\bibinfo  {journal} {Phys. Rev. Lett.}\ }\textbf {\bibinfo
  {volume} {74}},\ \bibinfo {pages} {4087} (\bibinfo {year}
  {1995})}\BibitemShut {NoStop}%
\bibitem [{\citenamefont {Cirac}\ and\ \citenamefont
  {Zoller}(1995)}]{Cirac1995}%
  \BibitemOpen
  \bibfield  {author} {\bibinfo {author} {\bibfnamefont {J.~I.}\ \bibnamefont
  {Cirac}}\ and\ \bibinfo {author} {\bibfnamefont {P.}~\bibnamefont {Zoller}},\
  }\href {\doibase 10.1103/PhysRevLett.74.4091} {\bibfield  {journal} {\bibinfo
   {journal} {Phys. Rev. Lett.}\ }\textbf {\bibinfo {volume} {74}},\ \bibinfo
  {pages} {4091} (\bibinfo {year} {1995})}\BibitemShut {NoStop}%
\bibitem [{\citenamefont {S\o{}rensen}\ and\ \citenamefont
  {M\o{}lmer}(1999)}]{Sorensen1999}%
  \BibitemOpen
  \bibfield  {author} {\bibinfo {author} {\bibfnamefont {A.}~\bibnamefont
  {S\o{}rensen}}\ and\ \bibinfo {author} {\bibfnamefont {K.}~\bibnamefont
  {M\o{}lmer}},\ }\href {\doibase 10.1103/PhysRevLett.82.1971} {\bibfield
  {journal} {\bibinfo  {journal} {Phys. Rev. Lett.}\ }\textbf {\bibinfo
  {volume} {82}},\ \bibinfo {pages} {1971} (\bibinfo {year}
  {1999})}\BibitemShut {NoStop}%
\bibitem [{\citenamefont {Ospelkaus}\ \emph {et~al.}(2008)\citenamefont
  {Ospelkaus}, \citenamefont {Langer}, \citenamefont {Amini}, \citenamefont
  {Brown}, \citenamefont {Leibfried},\ and\ \citenamefont
  {Wineland}}]{Ospelkaus2008}%
  \BibitemOpen
  \bibfield  {author} {\bibinfo {author} {\bibfnamefont {C.}~\bibnamefont
  {Ospelkaus}}, \bibinfo {author} {\bibfnamefont {C.~E.}\ \bibnamefont
  {Langer}}, \bibinfo {author} {\bibfnamefont {J.~M.}\ \bibnamefont {Amini}},
  \bibinfo {author} {\bibfnamefont {K.~R.}\ \bibnamefont {Brown}}, \bibinfo
  {author} {\bibfnamefont {D.}~\bibnamefont {Leibfried}}, \ and\ \bibinfo
  {author} {\bibfnamefont {D.~J.}\ \bibnamefont {Wineland}},\ }\href {\doibase
  10.1103/PhysRevLett.101.090502} {\bibfield  {journal} {\bibinfo  {journal}
  {Phys. Rev. Lett.}\ }\textbf {\bibinfo {volume} {101}},\ \bibinfo {pages}
  {090502} (\bibinfo {year} {2008})}\BibitemShut {NoStop}%
\bibitem [{\citenamefont {Schmidt-Kaler}\ \emph {et~al.}(2003)\citenamefont
  {Schmidt-Kaler}, \citenamefont {Haffner}, \citenamefont {Riebe},
  \citenamefont {Gulde}, \citenamefont {Lancaster}, \citenamefont {Deuschle},
  \citenamefont {Becher}, \citenamefont {Roos}, \citenamefont {Eschner},\ and\
  \citenamefont {Blatt}}]{SchmidtKaler2003}%
  \BibitemOpen
  \bibfield  {author} {\bibinfo {author} {\bibfnamefont {F.}~\bibnamefont
  {Schmidt-Kaler}}, \bibinfo {author} {\bibfnamefont {H.}~\bibnamefont
  {Haffner}}, \bibinfo {author} {\bibfnamefont {M.}~\bibnamefont {Riebe}},
  \bibinfo {author} {\bibfnamefont {S.}~\bibnamefont {Gulde}}, \bibinfo
  {author} {\bibfnamefont {G.~P.~T.}\ \bibnamefont {Lancaster}}, \bibinfo
  {author} {\bibfnamefont {T.}~\bibnamefont {Deuschle}}, \bibinfo {author}
  {\bibfnamefont {C.}~\bibnamefont {Becher}}, \bibinfo {author} {\bibfnamefont
  {C.~F.}\ \bibnamefont {Roos}}, \bibinfo {author} {\bibfnamefont
  {J.}~\bibnamefont {Eschner}}, \ and\ \bibinfo {author} {\bibfnamefont
  {R.}~\bibnamefont {Blatt}},\ }\href@noop {} {\bibfield  {journal} {\bibinfo
  {journal} {Nature}\ }\textbf {\bibinfo {volume} {422}},\ \bibinfo {pages}
  {408} (\bibinfo {year} {2003})}\BibitemShut {NoStop}%
\bibitem [{\citenamefont {Lanyon}\ \emph {et~al.}(2011)\citenamefont {Lanyon},
  \citenamefont {Hempel}, \citenamefont {Nigg}, \citenamefont {Mueller},
  \citenamefont {Gerritsma}, \citenamefont {Zaehringer}, \citenamefont
  {Schindler}, \citenamefont {Barreiro}, \citenamefont {Rambach}, \citenamefont
  {Kirchmair}, \citenamefont {Hennrich}, \citenamefont {Zoller}, \citenamefont
  {Blatt},\ and\ \citenamefont {Roos}}]{Lanyon2011}%
  \BibitemOpen
  \bibfield  {author} {\bibinfo {author} {\bibfnamefont {B.~P.}\ \bibnamefont
  {Lanyon}}, \bibinfo {author} {\bibfnamefont {C.}~\bibnamefont {Hempel}},
  \bibinfo {author} {\bibfnamefont {D.}~\bibnamefont {Nigg}}, \bibinfo {author}
  {\bibfnamefont {M.}~\bibnamefont {Mueller}}, \bibinfo {author} {\bibfnamefont
  {R.}~\bibnamefont {Gerritsma}}, \bibinfo {author} {\bibfnamefont
  {F.}~\bibnamefont {Zaehringer}}, \bibinfo {author} {\bibfnamefont
  {P.}~\bibnamefont {Schindler}}, \bibinfo {author} {\bibfnamefont {J.~T.}\
  \bibnamefont {Barreiro}}, \bibinfo {author} {\bibfnamefont {M.}~\bibnamefont
  {Rambach}}, \bibinfo {author} {\bibfnamefont {G.}~\bibnamefont {Kirchmair}},
  \bibinfo {author} {\bibfnamefont {M.}~\bibnamefont {Hennrich}}, \bibinfo
  {author} {\bibfnamefont {P.}~\bibnamefont {Zoller}}, \bibinfo {author}
  {\bibfnamefont {R.}~\bibnamefont {Blatt}}, \ and\ \bibinfo {author}
  {\bibfnamefont {C.~F.}\ \bibnamefont {Roos}},\ }\href {\doibase
  10.1126/science.1208001} {\bibfield  {journal} {\bibinfo  {journal}
  {Science}\ }\textbf {\bibinfo {volume} {333}},\ \bibinfo {pages} {57}
  (\bibinfo {year} {2011})}\BibitemShut {NoStop}%
\bibitem [{\citenamefont {Ospelkaus}\ \emph {et~al.}(2011)\citenamefont
  {Ospelkaus}, \citenamefont {Warring}, \citenamefont {Colombe}, \citenamefont
  {Brown}, \citenamefont {Amini}, \citenamefont {Leibfried},\ and\
  \citenamefont {Wineland}}]{Ospelkaus2011}%
  \BibitemOpen
  \bibfield  {author} {\bibinfo {author} {\bibfnamefont {C.}~\bibnamefont
  {Ospelkaus}}, \bibinfo {author} {\bibfnamefont {U.}~\bibnamefont {Warring}},
  \bibinfo {author} {\bibfnamefont {Y.}~\bibnamefont {Colombe}}, \bibinfo
  {author} {\bibfnamefont {K.~R.}\ \bibnamefont {Brown}}, \bibinfo {author}
  {\bibfnamefont {J.~M.}\ \bibnamefont {Amini}}, \bibinfo {author}
  {\bibfnamefont {D.}~\bibnamefont {Leibfried}}, \ and\ \bibinfo {author}
  {\bibfnamefont {D.~J.}\ \bibnamefont {Wineland}},\ }\href@noop {} {\bibfield
  {journal} {\bibinfo  {journal} {Nature}\ }\textbf {\bibinfo {volume} {476}},\
  \bibinfo {pages} {181} (\bibinfo {year} {2011})}\BibitemShut {NoStop}%
\bibitem [{\citenamefont {Blatt}\ and\ \citenamefont
  {Wineland}(2008)}]{Blatt2008}%
  \BibitemOpen
  \bibfield  {author} {\bibinfo {author} {\bibfnamefont {R.}~\bibnamefont
  {Blatt}}\ and\ \bibinfo {author} {\bibfnamefont {D.}~\bibnamefont
  {Wineland}},\ }\href@noop {} {\bibfield  {journal} {\bibinfo  {journal}
  {Nature}\ }\textbf {\bibinfo {volume} {453}},\ \bibinfo {pages} {1008}
  (\bibinfo {year} {2008})}\BibitemShut {NoStop}%
\bibitem [{\citenamefont {Hume}\ \emph {et~al.}(2007)\citenamefont {Hume},
  \citenamefont {Rosenband},\ and\ \citenamefont {Wineland}}]{Hume2007}%
  \BibitemOpen
  \bibfield  {author} {\bibinfo {author} {\bibfnamefont {D.~B.}\ \bibnamefont
  {Hume}}, \bibinfo {author} {\bibfnamefont {T.}~\bibnamefont {Rosenband}}, \
  and\ \bibinfo {author} {\bibfnamefont {D.~J.}\ \bibnamefont {Wineland}},\
  }\href {\doibase 10.1103/PhysRevLett.99.120502} {\bibfield  {journal}
  {\bibinfo  {journal} {Phys. Rev. Lett.}\ }\textbf {\bibinfo {volume} {99}},\
  \bibinfo {pages} {120502} (\bibinfo {year} {2007})}\BibitemShut {NoStop}%
\end{thebibliography}%


\end{document}